\documentclass[journal=jacsat,manuscript=article,amsmath,amssymb]{achemso}

\usepackage[version=3]{mhchem} 
\usepackage{braket}

\author{Menglin Huang}
\email{menglinhuang@fudan.edu.cn}
\affiliation{Key Laboratory of Computational Physical Sciences (MOE), Fudan University, Shanghai 200433, China}
\alsoaffiliation{School of Microelectronics, Fudan University, Shanghai 200433, China}
\author{Shanshan Wang}
\affiliation{Key Laboratory of Computational Physical Sciences (MOE), Fudan University, Shanghai 200433, China}
\author{Shiyou Chen}
\email{chensy@fudan.edu.cn}
\affiliation{Key Laboratory of Computational Physical Sciences (MOE), Fudan University, Shanghai 200433, China}
\alsoaffiliation{School of Microelectronics, Fudan University, Shanghai 200433, China}
\alsoaffiliation{Shanghai Qi Zhi Institute, Shanghai 200030, China}

\title[An \textsf{achemso} demo]
  {Metastability and anharmonicity enhance defect-assisted nonradiative recombination in low-symmetry semiconductors}

\abbreviations{IR,NMR,UV}
\keywords{American Chemical Society, \LaTeX}

\begin{document}


\begin{abstract}
Strong nonradiative recombination has been observed in quasi-one-dimensional antimony selenide, which runs counter to the simple intuition that claims high defect tolerance exists in semiconductors with antibonding state in the valence band and bonding state in the conduction band. Here we reveal such a defect intolerance actually stems from the richness of structural metastability and vibrational anharmonicity owing to the low-symmetry atomic structure. Taking the deep defect V$_{\rm Se}$ as a benchmark, we show the defect with its ground-state configuration alone does not act as a recombination center. Instead, we identify three different configurations with different formation energies, such richness of metastability offers a higher probability to accomplish a rapid recombination cycle. Another contributing factor is the anharmonicity in the potential energy surfaces that is caused by the large atomic relaxation, which elevates the total capture coefficient by 2-3 orders of magnitude compared with harmonic approximation. Therefore, the unique properties from both crystals and phonons in quasi-one-dimensional system enhance the nonradiative recombination, making the traditional intuition of defect tolerance invalid. These results highlight the importance of the correct identification of metastable defects and phonon anharmonicity in the nonradiative recombination in low-symmetry semiconductors.
\end{abstract}

\section{Introduction}
\par Defect tolerance in semiconductors is a concept that represents the magnitude of defect-assisted nonradiative recombination \cite{walsh2017instilling, zhang2020correctly, kang2017high}. Usually, defects with deep levels in the band gap have both decent electron and hole capture rates, so a high density of deep defects is an indication of rapid nonradiative recombination and low defect tolerance \cite{zhang2022defect}. In traditional III-V and II-VI semiconductors with bonding state mainly from anion p orbital in valence band maximum (VBM) and antibonding state from cation s orbital in conduction band minimum (CBM), the defect tolerance is usually very low. For instance, cadmium vacancy, tellurium interstitial, and tellurium-on-cadmium antisite in cadmium telluride (CdTe) are reported to have deep defect levels and large nonradiative recombination rates, so the carrier lifetime of CdTe can be limited in the order of nanosecond \cite{yang2016non,kavanagh2021rapid,kavanagh2022impact}. By contrast, high defect tolerance has been observed in semiconductors with an “inverted” band structure compared to the previous one \cite{zakutayev2014defect}, that is, the VBM is composed of antibonding state and the CBM is composed of bonding state, such as hybrid perovskites with s-p repulsion and many copper-based semiconductors with p-d repulsion in the VBM \cite{kang2017high,zakutayev2014defect,yin2014unusual}.

\par Such an empirical judgement has been a preliminary selection rule for the search of defect tolerant semiconductors for optoelectronic applications \cite{huang2022searching}. However, there still exist exceptions. Quasi-one-dimensional A$_2$B$_3$ (A = Sb, Bi; B = S, Se) are those that satisfy this requirement. To be specific, the lone pair $ns^2$ from A-site atoms strongly hybridizes with B-site p orbitals, thus forming antibonding states in the VBM, while the CBM is composed of the bonding state of A-site p and B-site p orbitals \cite{huang2019complicated}. Therefore, it was expected that the defects in A$_2$B$_3$ should have high tolerance to defect following the empirical rule. However, experimental measurements showed the defects are actually intolerant; for instance, deep-level defects in Sb$_2$Se$_3$ have high densities and large carrier capture cross sections, leading to short carrier lifetime, which indicates that the nonradiative recombination is strong \cite{chen2017characterization, wen2018vapor, hobson2020defect}. Theoretical calculations also found that various deep-level defects exist in the lattice \cite{huang2019complicated,savory2019complex,stoliaroff2020deciphering,huang2021more}, but the mechanism of nonradiative recombination caused by those defects is still not clear. The contradiction not only hinders the optimization of low-symmetry-semiconductor-based optoelectronic devices, but also challenges the common intuition of defect tolerance in semiconductors.

\par As a representative, the deep-level defects in Sb$_2$Se$_3$ crystals are dominant in 
 crystals; they were believed to be anion-cation antisites Sb$_{\rm Se}$ and Se$_{\rm Sb}$, and selenium vacancies V$_{\rm Se}$ according to first-principles calculations \cite{huang2021more,savory2019complex}. Due to the low-symmetry of the bulk structure that two Sb cations and three Se anions are inequivalent, see Figure S1(a), those defects on inequivalent sites have different transition levels, namely $\varepsilon$(0/$+$) and $\varepsilon$(0/$-$) levels of antisites (Sb$_{\rm Se}$ and Se$_{\rm Sb}$) and $\varepsilon$(0/$+$) and $\varepsilon$($+$/2$+$) levels of V$_{\rm Se}$, which provide various pathways for carrier nonradiative recombination, making the recombination mechanism in the entire system quite complicated. Among these defects, recent progress on defect correlation in Sb$_2$Se$_3$ reveals that the antisite defects Sb$_{\rm Se}$ and Se$_{\rm Sb}$ can be simultaneously eliminated by properly tuning the Se content, but the density of V$_{\rm Se}$ remains high regardless of growth conditions \cite{huang2021more, zhang2022competing}. Therefore, it is necessary to investigate the nonradiative recombination mechanism of V$_{\rm Se}$, and further find out why the defects in low-symmetry semiconductors does not follow the empirical rule of defect tolerance.

\par In this work, we establish a picture of defect chemistry shown in Figure 1 that the main contribution to the defect tolerance in low-symmetry semiconductors are the structural metastability and phonon anharmonicity, instead of the antibonding and bonding characters in the band edge in high-symmetry semiconductors. To be specific, by taking V$_{\rm Se}$ in Sb$_2$Se$_3$ an example, we find V$_{\rm Se}$ with its ground-state structure alone does not act as an effective recombination center, but the transition to diverse metastable structures increases the probability to finish a rapid recombination cycle. Furthermore, phonon anharmonicity during charge-state transition can be common, which significantly increases the vibronic overlap in carrier capture. Those observations can be common in quasi-one-dimensional semiconductors with low-symmetry structures, but are usually not observed in other s-p hybridized system, such as hybrid perovskites. Therefore, the empirical rule for evaluating the defect tolerance may not hold in low-symmetry semiconductors, and one has to explicitly consider the role of structural metastability and phonon anharmonicity of the defects in these systems.

\begin{figure}
\includegraphics[width=155mm]{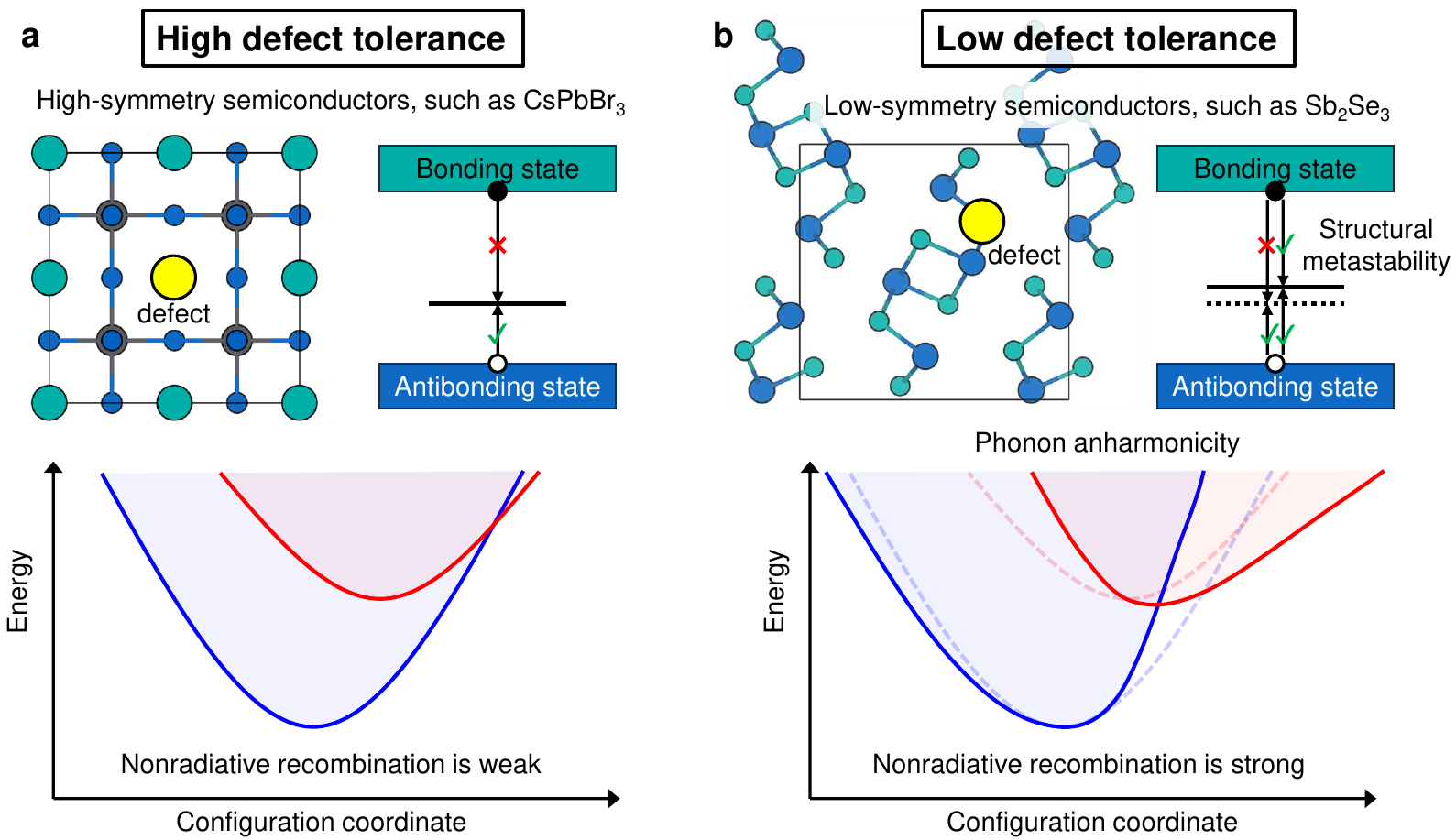}
\caption{Schematic illustration of defect tolerance in semiconductors with antibonding state in VBM and bonding state in CBM. (a) High defect tolerance is observed in high-symmetry semiconductors as expected. (b) Low defect tolerance is caused by structural metastability and phonon anharmonicity in low-symmetry semiconductors.}
\end{figure}

\section{Results and discussion}

\subsection{\label{sec:level2}Metastable defect configurations}

\par Due to the low symmetry of Sb$_2$Se$_3$ that inequivalent Sb and Se atoms exist in the crystal, the defect formations are complicated. For instance, there can be three different types of V$_{\rm Se}$ depending on the original atomic sites, namely V$_{\rm Se1}$, V$_{\rm Se2}$, V$_{\rm Se3}$ shown in Figure S1(b). They are all donor defects with negative-U $\varepsilon$(0/2$+$) charge-state transition levels within the band gap but with different formation energies. Among them, V$_{\rm Se1}$ has the lowest formation energy in neutral state but the highest formation energy in +1 and +2 charge states, causing the $\varepsilon$(0/2$+$) level close to the VBM. Our previous study demonstrated that only the density of V$_{\rm Se2}$ remains high \cite{huang2021more}, so we show in the article only the calculated results of V$_{\rm Se2}$ (which we label as V$_{\rm Se}$ below).

\begin{figure}
\includegraphics[width=130mm]{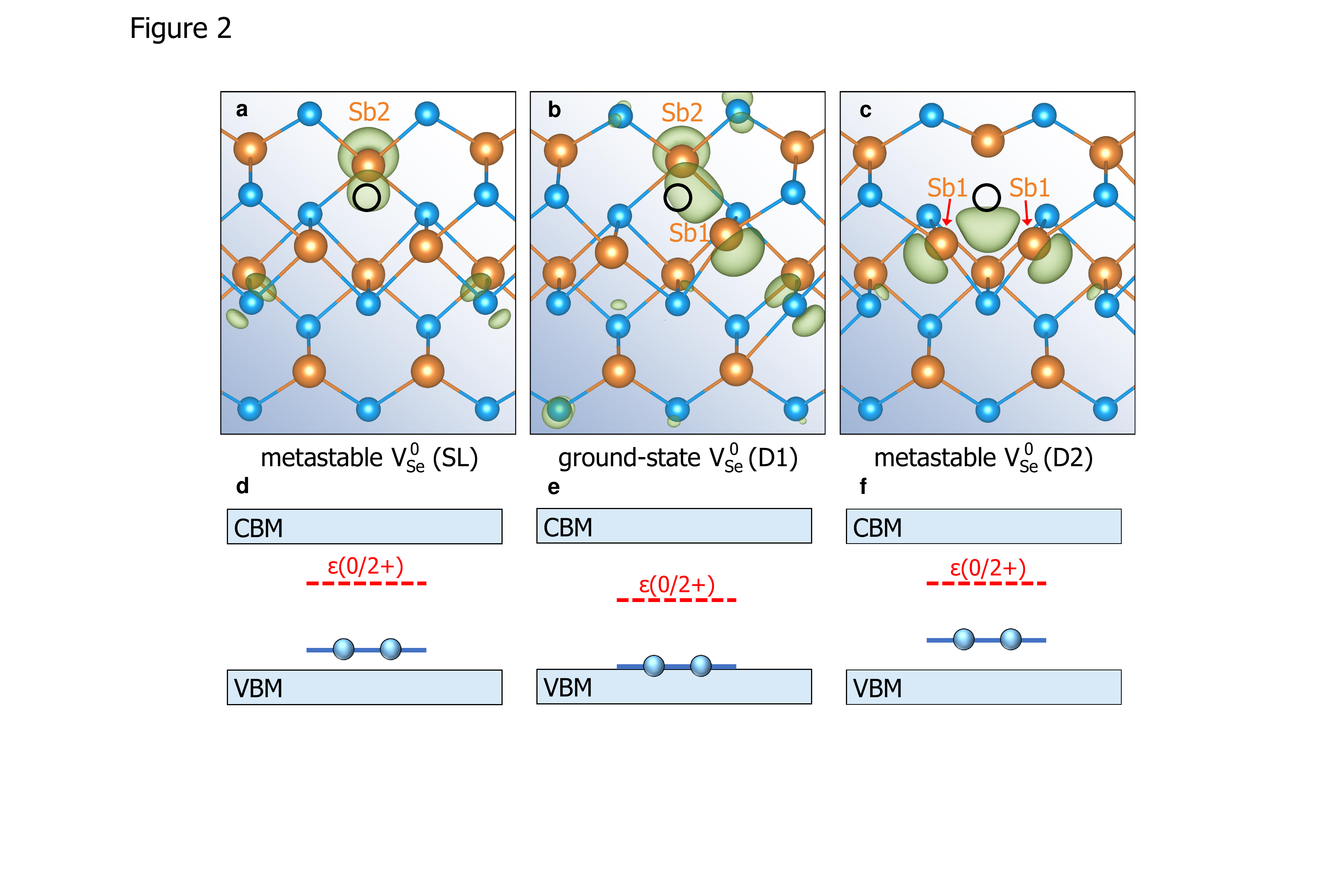}
\caption{Defect properties of V$_{\rm Se}$ in different configurations. (a-c) Local atomic geometries and charge-density distributions (in green) of V$_{\rm Se}$ in singly-localized (SL) structure, dimer (D1) structure, and another dimer (D2) structure in neutral state. (d-f) Comparison of single-particle levels (solid line occupied by electrons) and corresponding $\varepsilon$(0/2$+$) charge-state transition levels (dashed line) of the three possible configurations. The black circle shows the position of the Se vacancy.}
\end{figure}

\par Considering the low symmetry of quasi-one-dimensional structure, it is necessary to consider the possible metastable defect configurations in different charge states before we explicitly calculate the carrier capture of defect. Apart from the previously found structure of V$_{\rm Se}$ with charge density localized on one single Sb2 atom (labeled as SL in Figure 2(a)) in neutral state, which induces a $\varepsilon$(0/2$+$) charge-state transition level lying 0.78 eV above the VBM \cite{huang2019complicated}, the metastable searching algorithm implemented in DASP \cite{huang2022dasp} enables us to find two more structures with Sb dimer configurations. Figure 2(b) shows the first one (D1) with the dimerization of one Sb1 and one Sb2, forming a localized charge density connection between the two atoms; it is a bonding state of two Sb 5p orbitals. Figure 2(c) depicts another configuration (D2) with the dimerization of two Sb1 atoms. As we see in Table I, those metastable structures differ significantly from each other: they all have large structural differences $\Delta$\emph Q in configuration coordinate, indicating that the atomic relaxations between different structures are large. This is consistent with the theoretical results that claim the large void space between one-dimensional chains allows atomic distortions \cite{huang2019complicated}. The three possible configurations of V$_{\rm Se}$ in neutral state show different formation energies (Figure 3), in which D1 has the lowest while SL and D2 have similarly higher formation energies, so D1 configuration should be the ground-state structure of V$_{\rm Se}$, which is consistent with the recent theoretical result \cite{wang2023four}. The thermal transition barrier from the ground-state D1 structure to D2 calculated by transition-state theory is found to be 0.21 eV, while that from SL to D2 is only 0.11 eV, so D2 can theoretically co-exist at the room temperature.

\begin{table}
  \caption{Difference in configuration coordinate and formation energy between different metastable structures of V$_{\rm Se}$ in neutral state.}
  \label{tbl:example}
  \begin{tabular}{llll}
    \hline
Quantity & SL-D1 & SL-D2 & D1-D2\\
    \hline
$\Delta$\emph Q (amu$^{1/2}$ Å) & 8.67 & 15.48 & 15.28\\
$\Delta$\emph E$_f$ (eV) & 0.12 & -0.01 & $-$0.13\\
    \hline
  \end{tabular}
\end{table}

\par The different energetics of the three configurations are also reflected by the position of their single-particle defect levels found in the band gap, plotted by the solid line in Figure 2(d-f). Comparing the bonding states of D1 and D2 configurations that are composed of two Sb 5p orbitals, we note that the hybridization of D1 is stronger than that of D2. This can be reflected by the Sb-Sb distance in two configurations: in D1, the Sb1-Sb2 distance lowers from 4.09 Å to 3.08 Å, while the Sb2-Sb2 distance lowers from 3.96 Å to 3.11 Å. The larger atomic relaxation (larger bond length change during relaxation) and stronger hybridization (shorter Sb-Sb bond) cause the bonding state of D1 lying lower, consistent with the occupied defect level of D1 lying the lowest and its lowest defect formation energy. The corresponding $\varepsilon$(0/2$+$) transition levels are also compared (dashed line) in the band diagram to reflect the magnitude of lattice relaxation from 0/2+ transition.

\begin{figure}
\includegraphics[width=90mm]{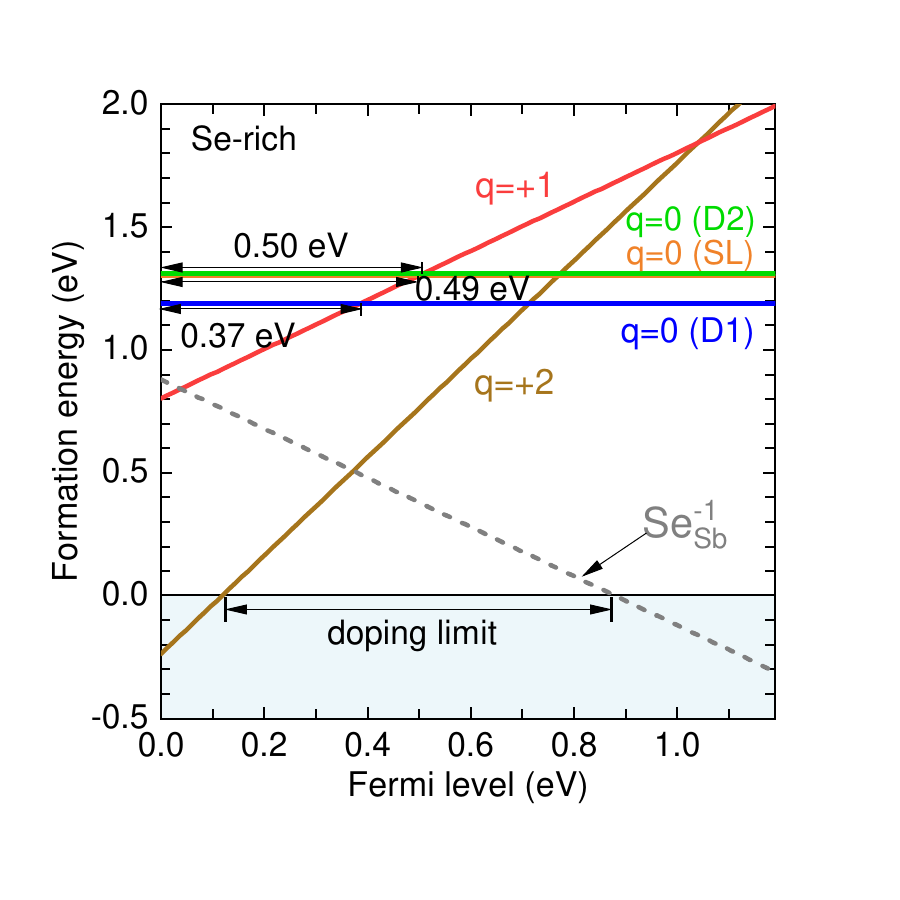}
\caption{Formation energies of V$_{\rm Se}$ with different configurations in different charge states as functions of Fermi level under Se-rich condition. The formation energy of Se$_{\rm Sb}^{-1}$ is also shown to imply the Fermi level pinnning. The Fermi level corresponding to the crossing of the formation energy line and the horizontal line is the doping limit.}
\end{figure}

\par Unlike neutral state, no extra configurations can be found for +1 and +2 charge states, whose charge densities are plotted in Figure S2 of Supplemental Material. Although the $\varepsilon$(0/2$+$) level roughly describes how charged defects affect the electrical properties of Sb$_2$Se$_3$, it does not reflect how the free carriers are captured by defects; that is, the carrier capture process only occurs at $\varepsilon$(0/+) and $\varepsilon$(+/2+) charge-state transition levels shown in Figure 3, due to the large Coulomb repulsion between two carriers during a direct 0/2$+$ transition. Specifically, the carrier capture at $\varepsilon$(0/$+$) has three different pathways but only one at $\varepsilon$($+$/2$+$). For an entire recombination cycle in the system, there can be totally four carrier capture coefficients, namely \emph C$_n^{2+}$ (electron capture by V$_{\rm Se}^{2+}$), \emph C$_n^{+}$ (electron capture by V$_{\rm Se}^{+}$), \emph C$_p^{+}$ (hole capture by V$_{\rm Se}^{+}$), and \emph C$_p^{0}$ (hole capture by V$_{\rm Se}^{0}$), which decide the total capture coefficient \emph C$_{\rm tot}$ according to Equation (14).

\subsection{\label{sec:level2}Carrier capture coefficients}

\begin{figure}
\includegraphics[width=160mm]{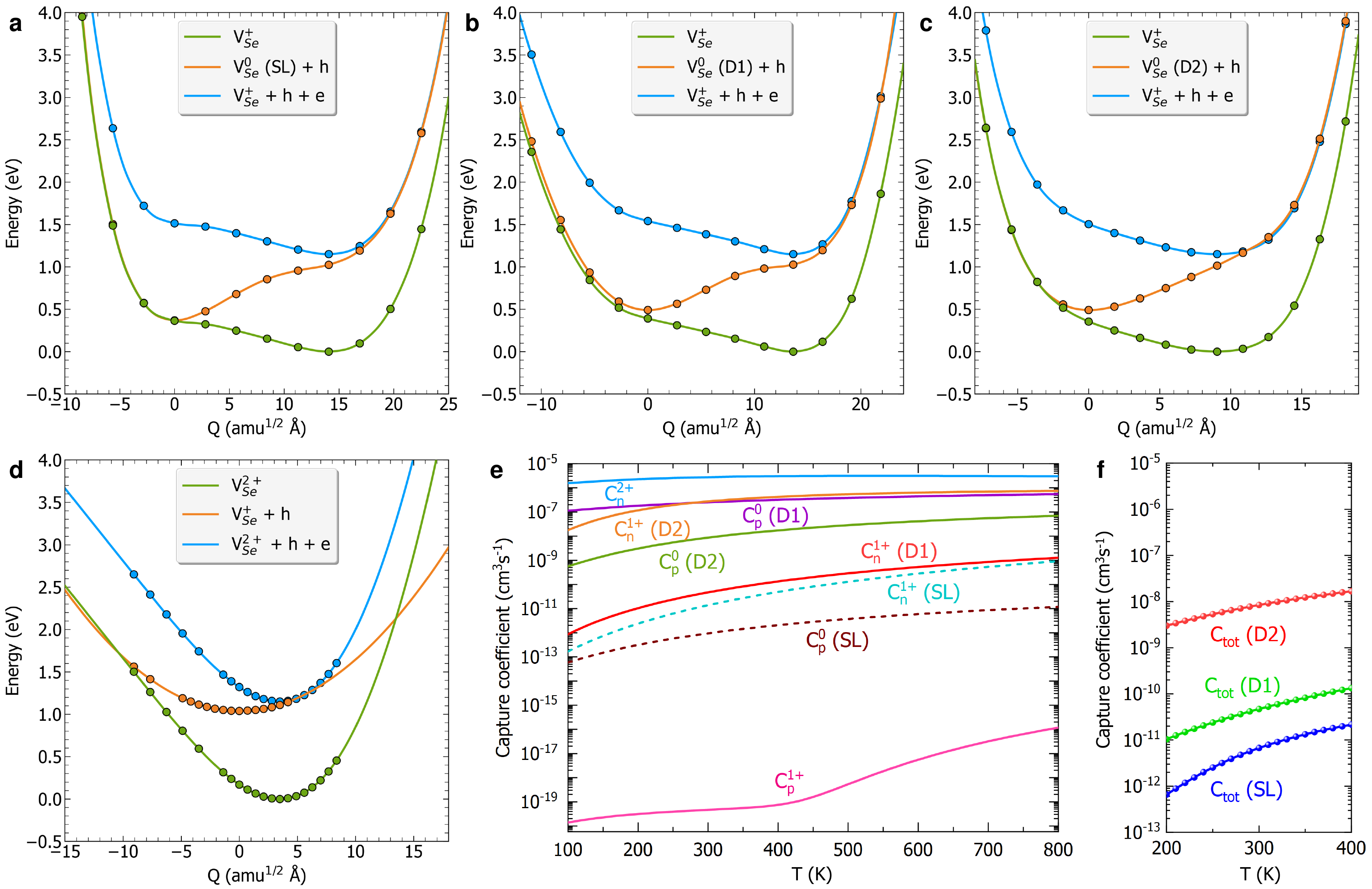}
\caption{Carrier capture process of V$_{\rm Se}$ in different charge states. (a-c) Configuration coordinate diagrams of carrier captures between q=0 and q=+1 charge states with different structures of neutral V$_{\rm Se}$. (d) Configuration coordinate diagram of carrier capture between q=+1 and q=+2 charge states. (e) Electron and hole carrier capture coefficients as functions of temperature. (f) Total carrier capture coefficients as functions of temperature.}
\end{figure}

\par Let us firstly focus on the carrier capture at $\varepsilon$(0/$+$). Since neutral V$_{\rm Se}$ has three different configurations, we show in Figure 4(a-c) the configuration coordinate diagrams for them. Figure 4(a) describes the carrier capture involving neutral SL structure. V$_{\rm Se}^{+}$ with a hole at the VBM and an electron at the CBM (blue curve) is the initial state, and it can capture the electron and become V$_{\rm Se}^{0}$ (SL), leading to a charge state transition. The classical barrier (the energy difference between the intersection point and the bottom of the blue curve) should be overcome to make the transition happen. Subsequently, V$_{\rm Se}^{0}$ (SL) as an intermediate state can also capture the hole at the VBM, becoming V$_{\rm Se}^{+}$ again. Such a process for hole capture will also be determined by the capture barrier between green and orange curves. For SL configuration both the electron and hole capture barriers are large. This would suggest that V$_{\rm Se}$ in SL configuration cannot be a possible recombination center due to the slow electron and hole capture rate. Following the similar analysis, Figure 4(b) indicates that the hole capture barrier of the ground-state D1 configuration is almost zero, significantly smaller than the large electron capture barrier. This shows the slower electron capture process may retard the entire recombination process as well. Thanks to the richness of metastability, Figure 4(c) shows that the D2 structure has minimal energy barriers both for electron and hole capture, implying the carrier capture process in the newly-found D2 configuration can be efficient. Those results suggest the transition to metastable structure is important for improving the carrier capture rate in semiconductors \cite{kavanagh2022impact,dou2023chemical}.

\par Figure 4(d) plots the configuration coordinate diagram for carrier capture at $\varepsilon$($+$/2$+$). As is reflected by the potential energy surfaces, the large hole capture barrier contrasting with the small electron capture barrier implies the hole capture can be very inefficient. However, according to Equation (14), such a process (\emph C$_p^{+}$, hole capture by V$_{\rm Se}^{+}$) is not the rate-limiting step for carrier recombination in a defect with neutral, +1 and +2 charge states, so its small value only have minor impacts on the total capture coefficient of V$_{\rm Se}$. Accurate calculation of carrier capture coefficient requires the exact calculation of vibrational wavefunction overlap and electron-phonon coupling matrix elements. We show in Figure 4(e) the carrier capture coefficients of the above processes. Following the same trend of our quantitative analysis, both the electron capture coefficient \emph C$_n^{+}$ and hole capture coefficient \emph C$_p^{0}$ of D2 (neutral) configuration are large, 2.8×10$^{-7}$ cm$^{3}$ s$^{-1}$ and 0.9×10$^{-8}$ cm$^{3}$ s$^{-1}$ at T=300 K, and the hole capture rate by V$_{\rm Se}^{+}$ is slow (\emph C$_p^{+}$ = 4.7×10$^{-20}$ cm$^{3}$ s$^{-1}$) at T=300 K. Figure 4(f) confirms that the total capture coefficient is not affected by the low value of \emph C$_p^{+}$, and the recombination involving D2 configuration is the most efficient one, contributing to a total capture coefficient \emph C$_{\rm tot}$ of 0.8×10$^{-8}$ cm$^{3}$ s$^{-1}$. 

\par It should be emphasized that the rapid nonradiative recombination of V$_{\rm Se}$ is contributed by the richness of structural metastability in quasi-one-dimensional crystals, because if only the ground-state structure configuration (D1) or the high-symmetry configuration (SL) are considered respectively, V$_{\rm Se}$ is not an effective recombination center. This can be viewed from the lower C$_{\rm tot}$ of 0.5×10$^{-10}$ cm$^{3}$ s$^{-1}$ for D1 and 0.9×10$^{-13}$ cm$^{3}$ s$^{-1}$ for SL, almost two and five orders of magnitude lower than that of D2 configuration. Supposing the defect density is about 10$^{16}$ cm$^{-3}$, the carrier recombination lifetime can be as large as 2 $\mu$s and 1 ms, much longer than the experimental measurement in the order of nanosecond \cite{chen2017characterization}.
\par One might wonder if the higher formation energies of the metastable defect configurations can lead to lower defect densities, which may play a role in the overall recombination rate. We would like to clarify that the total defect density of V$_{\rm Se}$ is not influenced by the different choices of neutral defect configuration. From Figure 3 we observe that the pinning Fermi level is calculated by the formation energies of V$_{\rm Se}^{2+}$ and Se$_{\rm Sb}^{-}$ \cite{huang2021more}. Once the Fermi level is determined by the intrinsic defects, the total density of V$_{\rm Se}$ is mainly contributed by that of V$_{\rm Se}^{2+}$, which does not have extra configurations with different formation energies, so it is expected that the density of V$_{\rm Se}$ is almost fixed when the configuration of V$_{\rm Se}^{0}$ changes. However, the doping limit denoted in Figure 3 shows if proper dopant is chosen, the Fermi level can be shifted up to 0.88 eV above the VBM and Sb$_2$Se$_3$ becomes n-type. In that case, metastability not only impacts the capture coefficient, but also changes the total defect densities that depend mainly on the neutral defect density.

\subsection{\label{sec:level2}Anharmonic potential energy surfaces}
\par Besides the role of metastable configuration of defects, another leading factor that causes the rapid nonradiative recombination by V$_{\rm Se}$ in quasi-one-dimensional system is the anharmonicity in the potential energy surfaces \cite{kim2019anharmonic,zhang2020correctly}. From the above analysis, we notice that the lattice relaxation in both charge-state transition and thermal structural transition (the transition between different configurations in neutral state) are quite large, with a similarly large configuration coordinate difference $\Delta$\emph Q (comparing Table I and Figure 4); this is in significant contrast with the defects in traditional materials. For instance, the relaxation of oxygen vacancies during charge-state transition is much smaller than that during thermal structural transition in SiO$_2$, so the harmonic approximation is applicable for the nonraditive multiphonon transition \cite{goes2018identification}. The unique characteristics of quasi-one-dimensional semiconductors leads to a different phenomenon in defect-assisted nonradiative recombination: anharmonicity might occur more likely due to the large deviation from the equilibrium structure.

\begin{figure}
\includegraphics[width=120mm]{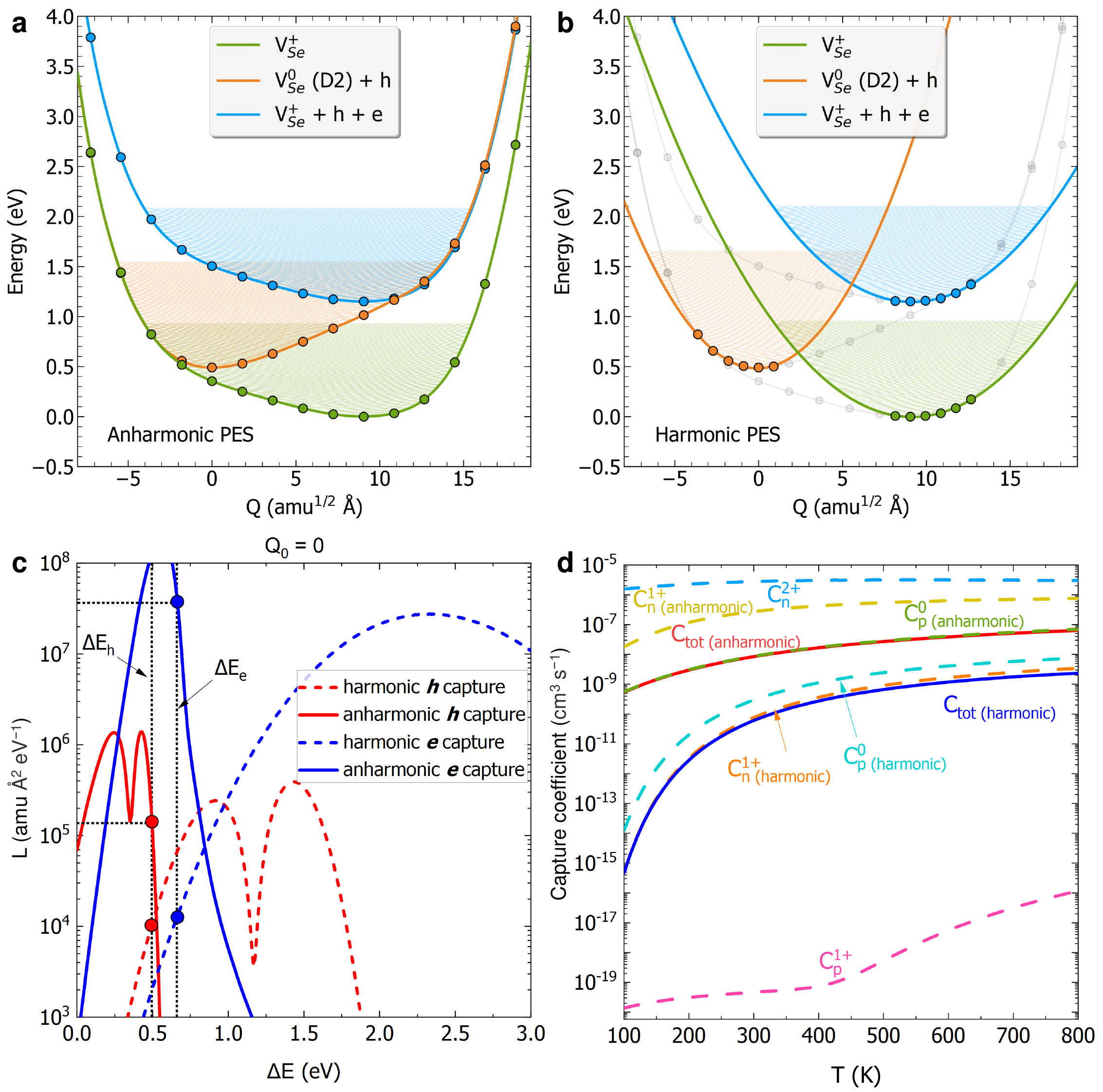}
\caption{Influence of anharmonicity on carrier capture. (a, b) Anharmonic and harmonic potential energy surfaces (PES) during the electron and hole capture at $\varepsilon$(0/$+$) transition level; the vibrational wavefunctions are obtained by solving one-dimensional Schrödinger equation and mapped in the configuration coordinate diagrams. (c) The intensity of lineshape function defined in Equation (1) as a function of carrier transition energy, in which \emph Q$_0$ is set to 0. (d) Comparison of carrier capture coefficients as functions of temperature.}
\end{figure}

\par The impact of including anharmonicity during carrier capture can be reflected by the vibrational wavefunctions plotted as the backgrounds in Figure 5(a,b), in which the wavefunctions and eigenvalues are obtained by directly solving time-independent one-dimensional Schrödinger equation along the anharmonic and harmonic potential energy surfaces, respectively. Apparently, using anharmonic potential energy surfaces (PES) results in very different wavefunctions as well as phonon overlaps. For instance, for the process of electron capture by V$_{\rm Se}^{+}$ (from blue to orange curves), anharmonic PES in Figure 5(a) shows a relatively weak electron-phonon coupling scenario that the intersection of the PES occurs at the same side at Q $>$ 10 amu$^{1/2}$ Å. By contrast, the harmonic PES picture shows the effective phonon energy of the final state (orange curve) is as small as 14.9 meV but with a large lattice relaxation energy of 2.2 eV, which corresponds to a Huang-Rhys factor of 148. Such a strong electron-phonon coupling shifts the intersection point into the middle of the two parabolic bottoms, which, however, increases the transition barrier. If we simply adopt the harmonic approximation to calculate the carrier capture of V$_{\rm Se}$, it is anticipated that the calculated results might be overestimated. To quantitatively assess such difference, a lineshape function is defined as,
\begin{equation}
L(\Delta E) = \sum_{\text{m}}p_m\sum_{\text{n}}\left| \bra{\chi_{im}}Q-Q_0\ket{\chi_{fn}} \right|^2 \delta (\Delta E+E_{im}-E_{fn}),    
\end{equation}
where $\delta$ function is replaced with a Gaussian, and the other parameters have been explained in Experimental Section. The lineshape function is a part of Equation (8) and a reflection of lattice relaxation and phonon energies. We plot in Figure 5(c) the lineshape function \emph L changing with the carrier transition energy $\Delta$\emph E both for anharmonic and harmonic carrier capture, where we choose \emph Q$_0$=0 in the configuration coordinate diagram both for electron and hole capture. It can be observed that a small increase in the transition energy $\Delta$\emph E may lead to a large shift of \emph L up to several orders of magnitude for anharmonic case, while the shift of \emph L is much more flat in harmonic case. For electron capture at $\Delta$\emph E=$\Delta$\emph E$_e$, \emph L is larger than 10$^7$ amu Å$^2$ eV$^{-1}$ when adopting anharmonic condition but lowers by three orders of magnitude if anharmonicity is neglected. The difference is one order of magnitude for hole capture. In Figure S3 of the Supplemental Material, we prove that such difference is not influenced by the choice of \emph Q$_0$, and even by the case where \emph Q $-$\emph Q$_0$ is pulled out from the integral. The lower value of lineshape function by using harmonic potential energy surfaces implies that the capture coefficient may be lowered as well. In Figure 5(d), we compare the calculated coefficients for those cases, and the results indicate that even if we have adopted the most efficient D2 configuration, the total capture coefficient \emph C$_{\rm tot}$ is 0.6×10$^{-11}$ cm$^{3}$ s$^{-1}$, nearly two orders of magnitude lower if the anharmonicity is not considered.

\par Considering the high density of V$_{\rm Se}$, the total capture coefficient \emph C$_{\rm tot}$ of 0.8×10$^{-8}$ cm$^{3}$ s$^{-1}$ will cause a relatively high value of Shockley-Read-Hall (SRH) recombination coefficient \emph A,
\begin{equation}
A=C_{tot}N_{D},
\end{equation}
where \emph N$_D$ is the summed density of V$_{\rm Se}$ in neutral, +1 and +2 charge states. When the density is in the order of 10$^{16}$ cm$^{-3}$  \cite{huang2021more}, the coefficient \emph A is as large as 0.8×10$^8$ s$^{-1}$, which corresponds to an SRH recombination lifetime of 13 ns. Such a result is in accord with the experimentally measured minority carrier lifetime for \emph p-type Sb$_2$Se$_3$ of 67 ns \cite{chen2017characterization}. By contrast, if we do not consider the anharmonicity, the resulting lifetime can be as long as 0.16 $\mu$s. To minimize the nonradiative recombination caused by V$_{\rm Se}$, it is necessary to lower its density by properly performing post selenization step with a careful control of Se content \cite{vidal2020efficient}, since excess Se would result in a higher density of V$_{\rm Se}$ and a reduction in carrier lifetime \cite{huang2021more}.

\ The defects in quasi-one-dimensional Sb$_2$Se$_3$ are complicated. It is likely that other defects may also have strong nonradiative recombination contributed by metastability and anharmonicity. There are also a number of lone-pair-based semiconductors with low-symmetry crystal structure with s-p hybridization in the VBM, such as Sb$_2$S$_3$, SbSeI, GeSe, SnO. When assessing the defect tolerance in those systems, one should keep in mind the picture in Figure 1 that both the metastable structures and phonon anharmonicity may contribute a more efficient nonradiative recombination cycle, so the simple intuition does not hold for those cases.

\section{Conclusions}

\par In summary, we found that the defect intolerance in quasi-one-dimensional semiconductor stems from the structural metastability and vibrational anharmonicity, both of which benefit from the low-symmetry crystals. Specifically, it is found V$_{\rm Se}$ in Sb$_2$Se$_3$ with its ground-state structure cannot act as an effective recombination center due to the slow electron capture during +/0 transition. However, the easy transition from stable to metastable structure increases the probability for efficient carrier capture. Furthermore, the anharmonicity in the potential energy surfaces can be common in such systems, which significantly lowers the transition barriers, so the nonradiative recombination becomes rapid. This study highlights the importance of metastable configuration and anharmonicity of defects in nonradiative recombination in low-symmetry semiconductors.

\section{Experimental}

\subsection{\label{sec:level2}Defect formation energy}

\par Calculations are performed based on density functional theory as implemented in the Vienna ab initio simulation package (VASP) \cite{kresse1993ab}. Hybrid functional of Heyd-Scuseria-Ernzerhof (HSE) \cite{heyd2003hybrid} form together with DFT-D3 dispersion correction \cite{grimme2010consistent} is adopted both for atomic relaxation and total energy calculation. The detailed parameters can be refered to Ref. \cite{huang2021more}.

\par The formation energy of selenium vacancies (V$_{\rm Se}$) in charge state \emph q can be calculated as \cite{freysoldt2014first},
\begin{equation}
\Delta E_f[\rm V_{\rm Se}^{\emph q}] = \emph E_{\rm tot}[\rm V_{\rm Se}^{\emph q}]-\emph E_{\rm tot}[\rm bulk]+\mu _{\rm Se}+\emph q \emph E_{\rm F}+\emph E_{\rm corr},
\end{equation}
where \emph E$_{\rm tot}$[V$_{\rm Se}^{q}$] and \emph E$_{\rm tot}$[bulk] are the total energies of supercell with and without a selenium vacancy, $\mu$$_{\rm Se}$ is the chemical potential of Se, \emph E$_{\rm F}$ is the Fermi level referenced to valence band maximum (VBM) calculated in bulk supercell at $\Gamma$ point, \emph E$_{\rm corr}$ is the correction that accounts for the finite-size effect \cite{freysoldt2009fully}.
\par When the defect formation energies in 0, + and 2+ charge states are calculated, the charge-state transition level can be calculated by,
\begin{equation}
\varepsilon(0/+)=\Delta E_f[{\rm V}_{\rm Se}^{0}; E_{\rm F}=0]-\Delta E_f[{\rm V}_{\rm Se}^{+}; E_{\rm F}=0],
\end{equation}
\begin{equation}
\varepsilon(+/2+)=\Delta E_f[{\rm V}_{\rm Se}^{+}; E_{\rm F}=0]-\Delta E_f[{\rm V}_{\rm Se}^{2+}; E_{\rm F}=0].
\end{equation}

\subsection{\label{sec:level2}Configuration coordinate diagram}
\par There should theoretically be 3N (N is the number of the atoms in the supercell) phonon modes in the system. In each mode \emph k, the displacement in configuration coordinate during charge-state transition can be expressed by \cite{alkauskas2014first},
\begin{equation}
\Delta Q_k=\sum_{\alpha}{\sqrt{m_\alpha}}\mu_k (\alpha)(R_{i;\alpha}-R_{f;\alpha}),
\end{equation}
where $\mu$$_k$($\alpha$) is the eigenvector of phonon mode \emph k, \emph m$_\alpha$ is the mass of the $\alpha$$^{th}$ atom in the supercell, \emph R$_{i;\alpha}$ and \emph R$_{f;\alpha}$ is the cartesian coordinate of the $\alpha$$^{th}$ atom in the supercells before and after charge-state transition, respectively.
\par We adopt a special vibrational mode to represent the phonons associated with the charge-state transition of defect, which is also known as the accepting mode \cite{schanovsky2011multiphonon,alkauskas2012first}. Such mode is defined by the difference between the cartesian coordinate in the supercells before and after charge-state transition. With the definition, the displacement can be expressed by \cite{alkauskas2012first,li2017large},
\begin{equation}
\Delta Q=\sqrt{\sum_{\alpha}{m_\alpha}{(R_{i;\alpha}-R_{f;\alpha})^2}}.
\end{equation}

\par By linearly interpolating the displacements using Equation (7) both for initial and final state structures, we can get the potential energy surfaces for them. For carrier capture by defect, the initial and final state potential energy surfaces are offset by $\Delta$\emph Q horizontally and by the energy of the transition vertically.

\subsection{\label{sec:level2}Nonradiative carrier capture}
\par We use static coupling theory based on Fermi's golden rule to calculate the carrier capture coefficient \cite{alkauskas2014nonrad},

\begin{equation}
C_n=fV\frac{2\pi}{\hbar}W_{if}^2\sum_{\text{m}}p_m\sum_{\text{n}}\left| \bra{\chi_{im}}Q-Q_0\ket{\chi_{fn}} \right|^2 \delta (\Delta E+E_{im}-E_{fn}),
\end{equation}

where \emph f is the Sommerfeld factor, $V$ is the supercell volume, \emph p$_m$ is the thermal occupation of the initial vibrational state \emph m, $\chi$$_{im}$ and $\chi$$_{fn}$ are the wavefunctions of the initial and final vibrational state \emph m and \emph n, \emph E$_{im}$ and \emph E$_{fn}$ are the corresponding eigenvalues. To account for the anharmonicity, these quantities are derived by directly solving the Schrödinger equation both for initial and final potential energy surfaces in the configuration coordinate diagram described above. $\Delta$\emph E is the carrier transition energy, which is correlated with the charge-state transition level of defects. \emph Q$_0$ refers to the equilibrium structure of initial or final state; it is necessary to carefully choose \emph Q$_0$ so that the localized defect state can be clearly identified within the band gap, instead of a resonant state with band edges. After \emph Q$_0$ is chosen, we use finite-difference method to calculate the electron-phonon coupling matrix element along the single phonon mode in the vicinity of \emph Q$_0$,
\begin{equation}
W_{if}=\bra{\psi_{i}}\frac{\partial H}{\partial Q}\ket{\psi_{f}}.
\end{equation}
\par The above calculations including the energetics of defects, the interpolation of potential energy surface, and the numerical calculation of capture coefficients are performed with the aid of Defect and dopant ab-initio simulation package (DASP) code \cite{huang2022dasp}.

\subsection{\label{sec:level2}Nonradiative recombination statistics}
\par Under steady-state illumination, the occupation at each defect transition level should not change \cite{sah1958electron}. In other words, there’s no charge-state transition between 0, +1 and +2 states of V$_{\rm Se}$. For instance, for charge-state transition at $\varepsilon$(+/2+), the electron capture rate (in unit of cm$^{-3}$ s$^{-1}$) by V$_{\rm Se}^{2+}$ must be equal to the hole capture rate by V$_{\rm Se}^{+}$,
\begin{equation}
R_n^{2+}=R_p^{+},
\end{equation}
considering that the equilibrium carrier density \emph n$_0$ and \emph p$_0$ of Sb$_2$Se$_3$ \cite{huang2021more} is lower than the typical photo-generated carrier density $\Delta$\emph n ($\Delta$\emph n=$\Delta$\emph p) \cite{wang2022effective}, the steady-state condition for carrier capture at $\varepsilon$(0/+) and $\varepsilon$(+/2+) can be respectively written as,
\begin{equation}
N^{2+}C_n^{2+}\Delta n=N^{+}C_p^{+}\Delta n,
\end{equation}
\begin{equation}
N^{+}C_n^{+}\Delta n=N^{0}C_p^{0}\Delta n.
\end{equation}

\par At steady-state condition, the overall recombination rate considering both processes at two levels is essentially the same as the total net electron or hole capture rate, 
\begin{equation}
(N^{2+}C_n^{2+}+N^{+}C_n^{+})\Delta n = (N^{+}C_p^{+}+N^{0}C_p^{0})\Delta n = NC_{tot}\Delta n, 
\end{equation}
combining Equations (11-13), the total carrier capture coefficient can be derived as \cite{alkauskas2016role,zhang2020iodine},
\begin{equation}
C_{tot}=\frac{C_n^{+}+C_p^{+}}{1+\frac{C_n^{+}}{C_p^{0}}+\frac{C_p^{+}}{C_n^{2+}}}.
\end{equation}

\begin{acknowledgement}

\par The authors thank Xinwei Wang and Prof. Aron Walsh for helpful discussion on the metastable structures of defects. This work was supported by National Key Research and Development Program of China (2019YFE0118100), China Postdoctoral Science Foundation Project (2022M720813), China National Postdoctoral Program for Innovative Talents (BX20230077), Ministry of Science and Technology of China (No. 2022YFA1402904), National Natural Science Foundation of China (12174060), Science and Technology Commission of Shanghai Municipality (Explorer project, 21TS1401000) and Project of MOE Innovation Platform.
\par M. Huang and S. Wang contributed equally to this work.

\end{acknowledgement}


\bibliography{acs-achemso}

\providecommand{\latin}[1]{#1}
\makeatletter
\providecommand{\doi}
  {\begingroup\let\do\@makeother\dospecials
  \catcode`\{=1 \catcode`\}=2 \doi@aux}
\providecommand{\doi@aux}[1]{\endgroup\texttt{#1}}
\makeatother
\providecommand*\mcitethebibliography{\thebibliography}
\csname @ifundefined\endcsname{endmcitethebibliography}  {\let\endmcitethebibliography\endthebibliography}{}
\begin{mcitethebibliography}{39}
\providecommand*\natexlab[1]{#1}
\providecommand*\mciteSetBstSublistMode[1]{}
\providecommand*\mciteSetBstMaxWidthForm[2]{}
\providecommand*\mciteBstWouldAddEndPuncttrue
  {\def\EndOfBibitem{\unskip.}}
\providecommand*\mciteBstWouldAddEndPunctfalse
  {\let\EndOfBibitem\relax}
\providecommand*\mciteSetBstMidEndSepPunct[3]{}
\providecommand*\mciteSetBstSublistLabelBeginEnd[3]{}
\providecommand*\EndOfBibitem{}
\mciteSetBstSublistMode{f}
\mciteSetBstMaxWidthForm{subitem}{(\alph{mcitesubitemcount})}
\mciteSetBstSublistLabelBeginEnd
  {\mcitemaxwidthsubitemform\space}
  {\relax}
  {\relax}

\bibitem[Walsh and Zunger(2017)Walsh, and Zunger]{walsh2017instilling}
Walsh,~A.; Zunger,~A. Instilling defect tolerance in new compounds. \emph{Nature Materials} \textbf{2017}, \emph{16}, 964--967\relax
\mciteBstWouldAddEndPuncttrue
\mciteSetBstMidEndSepPunct{\mcitedefaultmidpunct}
{\mcitedefaultendpunct}{\mcitedefaultseppunct}\relax
\EndOfBibitem
\bibitem[Zhang \latin{et~al.}(2020)Zhang, Turiansky, and Van~de Walle]{zhang2020correctly}
Zhang,~X.; Turiansky,~M.~E.; Van~de Walle,~C.~G. Correctly assessing defect tolerance in halide perovskites. \emph{The Journal of Physical Chemistry C} \textbf{2020}, \emph{124}, 6022--6027\relax
\mciteBstWouldAddEndPuncttrue
\mciteSetBstMidEndSepPunct{\mcitedefaultmidpunct}
{\mcitedefaultendpunct}{\mcitedefaultseppunct}\relax
\EndOfBibitem
\bibitem[Kang and Wang(2017)Kang, and Wang]{kang2017high}
Kang,~J.; Wang,~L.-W. High defect tolerance in lead halide perovskite CsPbBr$_3$. \emph{The Journal of Physical Chemistry Letters} \textbf{2017}, \emph{8}, 489--493\relax
\mciteBstWouldAddEndPuncttrue
\mciteSetBstMidEndSepPunct{\mcitedefaultmidpunct}
{\mcitedefaultendpunct}{\mcitedefaultseppunct}\relax
\EndOfBibitem
\bibitem[Zhang \latin{et~al.}(2022)Zhang, Turiansky, Shen, and Van~de Walle]{zhang2022defect}
Zhang,~X.; Turiansky,~M.~E.; Shen,~J.-X.; Van~de Walle,~C.~G. Defect tolerance in halide perovskites: A first-principles perspective. \emph{Journal of Applied Physics} \textbf{2022}, \emph{131}\relax
\mciteBstWouldAddEndPuncttrue
\mciteSetBstMidEndSepPunct{\mcitedefaultmidpunct}
{\mcitedefaultendpunct}{\mcitedefaultseppunct}\relax
\EndOfBibitem
\bibitem[Yang \latin{et~al.}(2016)Yang, Shi, Wang, and Wei]{yang2016non}
Yang,~J.-H.; Shi,~L.; Wang,~L.-W.; Wei,~S.-H. Non-radiative carrier recombination enhanced by two-level process: a first-principles study. \emph{Scientific Reports} \textbf{2016}, \emph{6}, 21712\relax
\mciteBstWouldAddEndPuncttrue
\mciteSetBstMidEndSepPunct{\mcitedefaultmidpunct}
{\mcitedefaultendpunct}{\mcitedefaultseppunct}\relax
\EndOfBibitem
\bibitem[Kavanagh \latin{et~al.}(2021)Kavanagh, Walsh, and Scanlon]{kavanagh2021rapid}
Kavanagh,~S.; Walsh,~A.; Scanlon,~D.~O. Rapid recombination by cadmium vacancies in CdTe. \emph{ACS Energy Letters} \textbf{2021}, \emph{6}, 1392--1398\relax
\mciteBstWouldAddEndPuncttrue
\mciteSetBstMidEndSepPunct{\mcitedefaultmidpunct}
{\mcitedefaultendpunct}{\mcitedefaultseppunct}\relax
\EndOfBibitem
\bibitem[Kavanagh \latin{et~al.}(2022)Kavanagh, Scanlon, Walsh, and Freysoldt]{kavanagh2022impact}
Kavanagh,~S.; Scanlon,~D.~O.; Walsh,~A.; Freysoldt,~C. Impact of metastable defect structures on carrier recombination in solar cells. \emph{Faraday Discussions} \textbf{2022}, \emph{239}, 339--356\relax
\mciteBstWouldAddEndPuncttrue
\mciteSetBstMidEndSepPunct{\mcitedefaultmidpunct}
{\mcitedefaultendpunct}{\mcitedefaultseppunct}\relax
\EndOfBibitem
\bibitem[Zakutayev \latin{et~al.}(2014)Zakutayev, Caskey, Fioretti, Ginley, Vidal, Stevanovic, Tea, and Lany]{zakutayev2014defect}
Zakutayev,~A.; Caskey,~C.~M.; Fioretti,~A.~N.; Ginley,~D.~S.; Vidal,~J.; Stevanovic,~V.; Tea,~E.; Lany,~S. Defect tolerant semiconductors for solar energy conversion. \emph{The Journal of Physical Chemistry Letters} \textbf{2014}, \emph{5}, 1117--1125\relax
\mciteBstWouldAddEndPuncttrue
\mciteSetBstMidEndSepPunct{\mcitedefaultmidpunct}
{\mcitedefaultendpunct}{\mcitedefaultseppunct}\relax
\EndOfBibitem
\bibitem[Yin \latin{et~al.}(2014)Yin, Shi, and Yan]{yin2014unusual}
Yin,~W.-J.; Shi,~T.; Yan,~Y. Unusual defect physics in CH$_3$NH$_3$PbI$_3$ perovskite solar cell absorber. \emph{Applied Physics Letters} \textbf{2014}, \emph{104}\relax
\mciteBstWouldAddEndPuncttrue
\mciteSetBstMidEndSepPunct{\mcitedefaultmidpunct}
{\mcitedefaultendpunct}{\mcitedefaultseppunct}\relax
\EndOfBibitem
\bibitem[Huang \latin{et~al.}(2022)Huang, Wang, Zhang, and Chen]{huang2022searching}
Huang,~M.; Wang,~S.; Zhang,~T.; Chen,~S. Searching for Band-Dispersive and Defect-Tolerant Semiconductors from Element Substitution in Topological Materials. \emph{Journal of the American Chemical Society} \textbf{2022}, \emph{144}, 4685--4694\relax
\mciteBstWouldAddEndPuncttrue
\mciteSetBstMidEndSepPunct{\mcitedefaultmidpunct}
{\mcitedefaultendpunct}{\mcitedefaultseppunct}\relax
\EndOfBibitem
\bibitem[Huang \latin{et~al.}(2019)Huang, Xu, Han, Tang, and Chen]{huang2019complicated}
Huang,~M.; Xu,~P.; Han,~D.; Tang,~J.; Chen,~S. Complicated and unconventional defect properties of the quasi-one-dimensional photovoltaic semiconductor Sb$_2$Se$_3$. \emph{ACS Applied Materials \& Interfaces} \textbf{2019}, \emph{11}, 15564--15572\relax
\mciteBstWouldAddEndPuncttrue
\mciteSetBstMidEndSepPunct{\mcitedefaultmidpunct}
{\mcitedefaultendpunct}{\mcitedefaultseppunct}\relax
\EndOfBibitem
\bibitem[Chen \latin{et~al.}(2017)Chen, Bobela, Yang, Lu, Zeng, Ge, Yang, Gao, Zhao, Beard, \latin{et~al.} others]{chen2017characterization}
Chen,~C.; Bobela,~D.~C.; Yang,~Y.; Lu,~S.; Zeng,~K.; Ge,~C.; Yang,~B.; Gao,~L.; Zhao,~Y.; Beard,~M.~C.; others Characterization of basic physical properties of Sb$_2$Se$_3$ and its relevance for photovoltaics. \emph{Frontiers of Optoelectronics} \textbf{2017}, \emph{10}, 18--30\relax
\mciteBstWouldAddEndPuncttrue
\mciteSetBstMidEndSepPunct{\mcitedefaultmidpunct}
{\mcitedefaultendpunct}{\mcitedefaultseppunct}\relax
\EndOfBibitem
\bibitem[Wen \latin{et~al.}(2018)Wen, Chen, Lu, Li, Kondrotas, Zhao, Chen, Gao, Wang, Zhang, \latin{et~al.} others]{wen2018vapor}
Wen,~X.; Chen,~C.; Lu,~S.; Li,~K.; Kondrotas,~R.; Zhao,~Y.; Chen,~W.; Gao,~L.; Wang,~C.; Zhang,~J.; others Vapor transport deposition of antimony selenide thin film solar cells with 7.6
\mciteBstWouldAddEndPuncttrue
\mciteSetBstMidEndSepPunct{\mcitedefaultmidpunct}
{\mcitedefaultendpunct}{\mcitedefaultseppunct}\relax
\EndOfBibitem
\bibitem[Hobson \latin{et~al.}(2020)Hobson, Phillips, Hutter, Durose, and Major]{hobson2020defect}
Hobson,~T.~D.; Phillips,~L.~J.; Hutter,~O.~S.; Durose,~K.; Major,~J.~D. Defect properties of Sb$_2$Se$_3$ thin film solar cells and bulk crystals. \emph{Applied Physics Letters} \textbf{2020}, \emph{116}\relax
\mciteBstWouldAddEndPuncttrue
\mciteSetBstMidEndSepPunct{\mcitedefaultmidpunct}
{\mcitedefaultendpunct}{\mcitedefaultseppunct}\relax
\EndOfBibitem
\bibitem[Savory and Scanlon(2019)Savory, and Scanlon]{savory2019complex}
Savory,~C.~N.; Scanlon,~D.~O. The complex defect chemistry of antimony selenide. \emph{Journal of Materials Chemistry A} \textbf{2019}, \emph{7}, 10739--10744\relax
\mciteBstWouldAddEndPuncttrue
\mciteSetBstMidEndSepPunct{\mcitedefaultmidpunct}
{\mcitedefaultendpunct}{\mcitedefaultseppunct}\relax
\EndOfBibitem
\bibitem[Stoliaroff \latin{et~al.}(2020)Stoliaroff, Lecomte, Rubel, Jobic, Zhang, Latouche, and Rocquefelte]{stoliaroff2020deciphering}
Stoliaroff,~A.; Lecomte,~A.; Rubel,~O.; Jobic,~S.; Zhang,~X.; Latouche,~C.; Rocquefelte,~X. Deciphering the role of key defects in Sb$_2$Se$_3$, a promising candidate for chalcogenide-based solar cells. \emph{ACS Applied Energy Materials} \textbf{2020}, \emph{3}, 2496--2509\relax
\mciteBstWouldAddEndPuncttrue
\mciteSetBstMidEndSepPunct{\mcitedefaultmidpunct}
{\mcitedefaultendpunct}{\mcitedefaultseppunct}\relax
\EndOfBibitem
\bibitem[Huang \latin{et~al.}(2021)Huang, Cai, Wang, Gong, Wei, and Chen]{huang2021more}
Huang,~M.; Cai,~Z.; Wang,~S.; Gong,~X.-G.; Wei,~S.-H.; Chen,~S. More Se vacancies in Sb$_2$Se$_3$ under Se-rich conditions: an abnormal behavior induced by defect-correlation in compensated compound semiconductors. \emph{Small} \textbf{2021}, \emph{17}, 2102429\relax
\mciteBstWouldAddEndPuncttrue
\mciteSetBstMidEndSepPunct{\mcitedefaultmidpunct}
{\mcitedefaultendpunct}{\mcitedefaultseppunct}\relax
\EndOfBibitem
\bibitem[Zhang and Qian(2022)Zhang, and Qian]{zhang2022competing}
Zhang,~B.; Qian,~X. Competing superior electronic structure and complex defect chemistry in quasi-one-dimensional antimony chalcogenide photovoltaic absorbers. \emph{ACS Applied Energy Materials} \textbf{2022}, \emph{5}, 492--502\relax
\mciteBstWouldAddEndPuncttrue
\mciteSetBstMidEndSepPunct{\mcitedefaultmidpunct}
{\mcitedefaultendpunct}{\mcitedefaultseppunct}\relax
\EndOfBibitem
\bibitem[Huang \latin{et~al.}(2022)Huang, Zheng, Dai, Guo, Wang, Jiang, Wei, and Chen]{huang2022dasp}
Huang,~M.; Zheng,~Z.; Dai,~Z.; Guo,~X.; Wang,~S.; Jiang,~L.; Wei,~J.; Chen,~S. DASP: Defect and dopant ab-initio simulation package. \emph{Journal of Semiconductors} \textbf{2022}, \emph{43}, 042101\relax
\mciteBstWouldAddEndPuncttrue
\mciteSetBstMidEndSepPunct{\mcitedefaultmidpunct}
{\mcitedefaultendpunct}{\mcitedefaultseppunct}\relax
\EndOfBibitem
\bibitem[Wang \latin{et~al.}(2023)Wang, Kavanagh, Scanlon, and Walsh]{wang2023four}
Wang,~X.; Kavanagh,~S.~R.; Scanlon,~D.~O.; Walsh,~A. Four-electron negative-$U$ vacancy defects in antimony selenide. \emph{Physical Review B} \textbf{2023}, \emph{108}, 134102\relax
\mciteBstWouldAddEndPuncttrue
\mciteSetBstMidEndSepPunct{\mcitedefaultmidpunct}
{\mcitedefaultendpunct}{\mcitedefaultseppunct}\relax
\EndOfBibitem
\bibitem[Dou \latin{et~al.}(2023)Dou, Falletta, Neugebauer, Freysoldt, Zhang, and Wei]{dou2023chemical}
Dou,~B.; Falletta,~S.; Neugebauer,~J.; Freysoldt,~C.; Zhang,~X.; Wei,~S.-H. Chemical Trend of Nonradiative Recombination in Cu(In,Ga)Se$_2$ Alloys. \emph{Physical Review Applied} \textbf{2023}, \emph{19}, 054054\relax
\mciteBstWouldAddEndPuncttrue
\mciteSetBstMidEndSepPunct{\mcitedefaultmidpunct}
{\mcitedefaultendpunct}{\mcitedefaultseppunct}\relax
\EndOfBibitem
\bibitem[Kim \latin{et~al.}(2019)Kim, Hood, and Walsh]{kim2019anharmonic}
Kim,~S.; Hood,~S.~N.; Walsh,~A. Anharmonic lattice relaxation during nonradiative carrier capture. \emph{Physical Review B} \textbf{2019}, \emph{100}, 041202\relax
\mciteBstWouldAddEndPuncttrue
\mciteSetBstMidEndSepPunct{\mcitedefaultmidpunct}
{\mcitedefaultendpunct}{\mcitedefaultseppunct}\relax
\EndOfBibitem
\bibitem[Goes \latin{et~al.}(2018)Goes, Wimmer, El-Sayed, Rzepa, Jech, Shluger, and Grasser]{goes2018identification}
Goes,~W.; Wimmer,~Y.; El-Sayed,~A.-M.; Rzepa,~G.; Jech,~M.; Shluger,~A.~L.; Grasser,~T. Identification of oxide defects in semiconductor devices: A systematic approach linking DFT to rate equations and experimental evidence. \emph{Microelectronics Reliability} \textbf{2018}, \emph{87}, 286--320\relax
\mciteBstWouldAddEndPuncttrue
\mciteSetBstMidEndSepPunct{\mcitedefaultmidpunct}
{\mcitedefaultendpunct}{\mcitedefaultseppunct}\relax
\EndOfBibitem
\bibitem[Vidal-Fuentes \latin{et~al.}(2020)Vidal-Fuentes, Placidi, S{\'a}nchez, Becerril-Romero, Andrade-Arvizu, Jehl, P{\'e}rez-Rodr{\'\i}guez, Izquierdo-Roca, and Saucedo]{vidal2020efficient}
Vidal-Fuentes,~P.; Placidi,~M.; S{\'a}nchez,~Y.; Becerril-Romero,~I.; Andrade-Arvizu,~J.; Jehl,~Z.; P{\'e}rez-Rodr{\'\i}guez,~A.; Izquierdo-Roca,~V.; Saucedo,~E. Efficient Se-Rich Sb$_2$Se$_3$/CdS Planar Heterojunction Solar Cells by Sequential Processing: Control and Influence of Se Content. \emph{Solar RRL} \textbf{2020}, \emph{4}, 2000141\relax
\mciteBstWouldAddEndPuncttrue
\mciteSetBstMidEndSepPunct{\mcitedefaultmidpunct}
{\mcitedefaultendpunct}{\mcitedefaultseppunct}\relax
\EndOfBibitem
\bibitem[Kresse and Hafner(1993)Kresse, and Hafner]{kresse1993ab}
Kresse,~G.; Hafner,~J. Ab initio molecular dynamics for liquid metals. \emph{Physical Review B} \textbf{1993}, \emph{47}, 558\relax
\mciteBstWouldAddEndPuncttrue
\mciteSetBstMidEndSepPunct{\mcitedefaultmidpunct}
{\mcitedefaultendpunct}{\mcitedefaultseppunct}\relax
\EndOfBibitem
\bibitem[Heyd \latin{et~al.}(2003)Heyd, Scuseria, and Ernzerhof]{heyd2003hybrid}
Heyd,~J.; Scuseria,~G.~E.; Ernzerhof,~M. Hybrid functionals based on a screened Coulomb potential. \emph{The Journal of Chemical Physics} \textbf{2003}, \emph{118}, 8207--8215\relax
\mciteBstWouldAddEndPuncttrue
\mciteSetBstMidEndSepPunct{\mcitedefaultmidpunct}
{\mcitedefaultendpunct}{\mcitedefaultseppunct}\relax
\EndOfBibitem
\bibitem[Grimme \latin{et~al.}(2010)Grimme, Antony, Ehrlich, and Krieg]{grimme2010consistent}
Grimme,~S.; Antony,~J.; Ehrlich,~S.; Krieg,~H. A consistent and accurate ab initio parametrization of density functional dispersion correction (DFT-D) for the 94 elements H-Pu. \emph{The Journal of Chemical Physics} \textbf{2010}, \emph{132}\relax
\mciteBstWouldAddEndPuncttrue
\mciteSetBstMidEndSepPunct{\mcitedefaultmidpunct}
{\mcitedefaultendpunct}{\mcitedefaultseppunct}\relax
\EndOfBibitem
\bibitem[Freysoldt \latin{et~al.}(2014)Freysoldt, Grabowski, Hickel, Neugebauer, Kresse, Janotti, and Van~de Walle]{freysoldt2014first}
Freysoldt,~C.; Grabowski,~B.; Hickel,~T.; Neugebauer,~J.; Kresse,~G.; Janotti,~A.; Van~de Walle,~C.~G. First-principles calculations for point defects in solids. \emph{Reviews of Modern Physics} \textbf{2014}, \emph{86}, 253\relax
\mciteBstWouldAddEndPuncttrue
\mciteSetBstMidEndSepPunct{\mcitedefaultmidpunct}
{\mcitedefaultendpunct}{\mcitedefaultseppunct}\relax
\EndOfBibitem
\bibitem[Freysoldt \latin{et~al.}(2009)Freysoldt, Neugebauer, and Van~de Walle]{freysoldt2009fully}
Freysoldt,~C.; Neugebauer,~J.; Van~de Walle,~C.~G. Fully ab initio finite-size corrections for charged-defect supercell calculations. \emph{Physical Review Letters} \textbf{2009}, \emph{102}, 016402\relax
\mciteBstWouldAddEndPuncttrue
\mciteSetBstMidEndSepPunct{\mcitedefaultmidpunct}
{\mcitedefaultendpunct}{\mcitedefaultseppunct}\relax
\EndOfBibitem
\bibitem[Alkauskas \latin{et~al.}(2014)Alkauskas, Buckley, Awschalom, and Van~de Walle]{alkauskas2014first}
Alkauskas,~A.; Buckley,~B.~B.; Awschalom,~D.~D.; Van~de Walle,~C.~G. First-principles theory of the luminescence lineshape for the triplet transition in diamond NV centres. \emph{New Journal of Physics} \textbf{2014}, \emph{16}, 073026\relax
\mciteBstWouldAddEndPuncttrue
\mciteSetBstMidEndSepPunct{\mcitedefaultmidpunct}
{\mcitedefaultendpunct}{\mcitedefaultseppunct}\relax
\EndOfBibitem
\bibitem[Schanovsky \latin{et~al.}(2011)Schanovsky, G{\"o}s, and Grasser]{schanovsky2011multiphonon}
Schanovsky,~F.; G{\"o}s,~W.; Grasser,~T. Multiphonon hole trapping from first principles. \emph{Journal of Vacuum Science \& Technology B} \textbf{2011}, \emph{29}\relax
\mciteBstWouldAddEndPuncttrue
\mciteSetBstMidEndSepPunct{\mcitedefaultmidpunct}
{\mcitedefaultendpunct}{\mcitedefaultseppunct}\relax
\EndOfBibitem
\bibitem[Alkauskas \latin{et~al.}(2012)Alkauskas, Lyons, Steiauf, and Van~de Walle]{alkauskas2012first}
Alkauskas,~A.; Lyons,~J.~L.; Steiauf,~D.; Van~de Walle,~C.~G. First-principles calculations of luminescence spectrum line shapes for defects in semiconductors: the example of GaN and ZnO. \emph{Physical Review Letters} \textbf{2012}, \emph{109}, 267401\relax
\mciteBstWouldAddEndPuncttrue
\mciteSetBstMidEndSepPunct{\mcitedefaultmidpunct}
{\mcitedefaultendpunct}{\mcitedefaultseppunct}\relax
\EndOfBibitem
\bibitem[Li \latin{et~al.}(2017)Li, Zhu, Zhang, Yuan, Chen, and Gong]{li2017large}
Li,~J.; Zhu,~H.-F.; Zhang,~Y.-Y.; Yuan,~Z.-K.; Chen,~S.; Gong,~X.-G. Large carrier-capture rate of $\mathrm{P}{\mathrm{b}}_{\mathrm{I}}$ antisite in $\mathrm{C}{\mathrm{H}}_{3}\mathrm{N}{\mathrm{H}}_{3}\mathrm{Pb}{\mathrm{I}}_{3}$ induced by heavy atoms and soft phonon modes. \emph{Physical Review B} \textbf{2017}, \emph{96}, 104103\relax
\mciteBstWouldAddEndPuncttrue
\mciteSetBstMidEndSepPunct{\mcitedefaultmidpunct}
{\mcitedefaultendpunct}{\mcitedefaultseppunct}\relax
\EndOfBibitem
\bibitem[Alkauskas \latin{et~al.}(2014)Alkauskas, Yan, and Van~de Walle]{alkauskas2014nonrad}
Alkauskas,~A.; Yan,~Q.; Van~de Walle,~C.~G. First-principles theory of nonradiative carrier capture via multiphonon emission. \emph{Physical Review B} \textbf{2014}, \emph{90}, 075202\relax
\mciteBstWouldAddEndPuncttrue
\mciteSetBstMidEndSepPunct{\mcitedefaultmidpunct}
{\mcitedefaultendpunct}{\mcitedefaultseppunct}\relax
\EndOfBibitem
\bibitem[Sah and Shockley(1958)Sah, and Shockley]{sah1958electron}
Sah,~C.-T.; Shockley,~W. Electron-hole recombination statistics in semiconductors through flaws with many charge conditions. \emph{Physical Review} \textbf{1958}, \emph{109}, 1103\relax
\mciteBstWouldAddEndPuncttrue
\mciteSetBstMidEndSepPunct{\mcitedefaultmidpunct}
{\mcitedefaultendpunct}{\mcitedefaultseppunct}\relax
\EndOfBibitem
\bibitem[Wang \latin{et~al.}(2022)Wang, Huang, Wu, Chu, Zhao, Walsh, Gong, Wei, and Chen]{wang2022effective}
Wang,~S.; Huang,~M.; Wu,~Y.-N.; Chu,~W.; Zhao,~J.; Walsh,~A.; Gong,~X.-G.; Wei,~S.-H.; Chen,~S. Effective lifetime of non-equilibrium carriers in semiconductors from non-adiabatic molecular dynamics simulations. \emph{Nature Computational Science} \textbf{2022}, \emph{2}, 486--493\relax
\mciteBstWouldAddEndPuncttrue
\mciteSetBstMidEndSepPunct{\mcitedefaultmidpunct}
{\mcitedefaultendpunct}{\mcitedefaultseppunct}\relax
\EndOfBibitem
\bibitem[Alkauskas \latin{et~al.}(2016)Alkauskas, Dreyer, Lyons, and Van~de Walle]{alkauskas2016role}
Alkauskas,~A.; Dreyer,~C.~E.; Lyons,~J.~L.; Van~de Walle,~C.~G. Role of excited states in Shockley-Read-Hall recombination in wide-band-gap semiconductors. \emph{Physical Review B} \textbf{2016}, \emph{93}, 201304\relax
\mciteBstWouldAddEndPuncttrue
\mciteSetBstMidEndSepPunct{\mcitedefaultmidpunct}
{\mcitedefaultendpunct}{\mcitedefaultseppunct}\relax
\EndOfBibitem
\bibitem[Zhang \latin{et~al.}(2020)Zhang, Turiansky, Shen, and Van~de Walle]{zhang2020iodine}
Zhang,~X.; Turiansky,~M.~E.; Shen,~J.-X.; Van~de Walle,~C.~G. Iodine interstitials as a cause of nonradiative recombination in hybrid perovskites. \emph{Physical Review B} \textbf{2020}, \emph{101}, 140101\relax
\mciteBstWouldAddEndPuncttrue
\mciteSetBstMidEndSepPunct{\mcitedefaultmidpunct}
{\mcitedefaultendpunct}{\mcitedefaultseppunct}\relax
\EndOfBibitem
\end{mcitethebibliography}

\end{document}